\documentclass[aps,prl,amsmath,twocolumn,groupedaddress,letterpaper,floatfix]{revtex4}
\usepackage{graphicx,color}
\usepackage{verbatim}
\usepackage{amssymb}   % for math
\usepackage{amsmath}
\usepackage{amsfonts}
\usepackage{mathdots}
\usepackage{hyperref}
\usepackage{epsfig}
\usepackage{braket}
\usepackage{bm}
\begin{document}
	\title{Nonlinear Hall Effects in Strained Twisted Bilayer WSe$_2$}
\author{Jin-Xin Hu$^{1}$}
\author{Cheng-Ping Zhang$^{1}$}
\author{Ying-Ming Xie$^{1}$}
\author{K. T. Law$^{1}$} \thanks{Correspondence author: phlaw@ust.hk}
\affiliation{Department of Physics, Hong Kong University of Science and Technology, Clear Water Bay, Hong Kong, China} 	
	\date{\today}
	\begin{abstract}
Recently, it has been pointed out that the twisting of bilayer WSe$_2$ would generate topologically non-trivial flat bands near the Fermi energy. In this work, we show that twisted bilayer WSe$_2$ (tWSe$_2$) with uniaxial strain exhibits a large nonlinear Hall (NLH) response due to the non-trivial Berry curvatures of the flat bands. Moreover, the NLH effect is greatly enhanced near the topological phase transition point which can be tuned by a vertical displacement field. Importantly, the nonlinear Hall signal changes sign across the topological phase transition point and provides a way to identify the topological phase transition and probe the topological properties of the flat bands. The strong enhancement and high tunability of the NLH effect near the topological phase transition point renders tWSe$_2$ and related moire materials available platforms for rectification and second harmonic generations. \end{abstract}
\pacs{}
\maketitle
\emph{Introduction.}---The study of long-period moir\'{e} superlattices formed in van der Waals heterostructures has emerged as a central topic in condensed matter physics\cite{Bistritzer}. After the observation of correlated insulator and superconductivity in twisted bilayer graphene(TBG) with flat bands \cite{cao2018correlated,cao2018unconventional,lu2019superconductors,yankowitz2019tuning,po2018origin,isobe2018unconventional,wu2018theory,xie2020nature,liu2019pseudo,koshino2018maximally}, it was proposed that there are moir\'{e}-mediated flat bands in twisted transition metal dichalcogenide heterobilayers and homobilayers \cite{wu2018hubbard,wu2019topological}. Recently, correlated insulating phases and possible superconductivity signatures were discovered in twisted bilayer WSe$_2$ at twist angles between $4^\circ$ to $5^\circ$\cite{wang2020correlated} and twisted double-bilayer WSe$_2$\cite{an2020interaction}.

Besides the correlated phases, it was shown that the flat bands of the moire superlattices also exhibit non-trivial topological properties. For example, at 3/4 filling in hBN-aligned TBG, the degeneracy of the bands with non-trivial Chern numbers is lifted by electron-electron interactions and results in quantum anomalous Hall states which were observed recently \cite{sharpe2019emergent,serlin2020intrinsic,he2020giant}. These observations clearly demonstrate that the topological properties of the flat bands also have important consequences on the nature of the correlated phases.

Similar to TBG, it was pointed out that flat bands with non-trivial Chern numbers can be generated in twisted bilayer WSe$_2$ (tWSe$_2$) and a couple of topological insulating phases were predicted at a wide range of twist angles\cite{wu2019topological}. As in the case of TBG, the topology of the bands would affect the nature of the insulating phase when the bands are half filled and the interaction effects are strong. In this work, we propose that the nonlinear response in electric field can be used to unveil the topological properties of the flat bands and identify the topological phase transition point through the measurements of the nonlinear Hall (NLH) effect.

The NLH effect is a fascinating phenomenon recently proposed by Sodemann and Fu\cite{sodemann2015quantum}, and experimentally observed in bilayer and multilayer WTe$_2$\cite{ma2019observation,kang2019nonlinear}. It is the generation of a transverse DC current and a transverse voltage with frequency $2\omega$ when an AC current of frequency $\omega$ is applied, which has potential applications in rectifications and second harmonic generations. The effect originates from the non-vanishing dipole moment of the Berry curvature of the bands which characterizes the second-order nonlinear hall susceptibility. In pristine tWSe$_2$, regardless of how large the Berry curvatures of the bands are, the three fold rotational symmetry forces the Berry curvature dipole to be zero and the NLH effect vanishes. However, we demonstrate here that a small strain breaks the three-fold rotational symmetry and generates a large Berry curvature dipole. Importantly, these symmetry breaking strain effects have been observed recently in experiments\cite{zhang2020flat}. Moreover, the Berry curvature dipole is strongly enhanced and has opposite signs across the topological phase transition point when the top two valence bands touch each other and exchange Berry curvatures\cite{facio2018strongly}. As the topological phase transition can be tuned by a vertical displacement field, the measurement of the NLH effect can serve as a probe of the topological phase transition in  tWSe$_2$. Moreover, the strong enhancement and the highly tunability of the NLH effect by a displacement field renders tWSe$_2$ and related moire materials available platforms for rectification and second harmonic generations applications.

The rest of our paper is organized as follows. First, we present the continuum model of tWSe$_2$ which takes into account a uniaxial strain in the bottom layer induced by substrate or external modulation, which breaks the three-fold rotational symmetry such that nonzero Berry curvature dipole can be created. Second, we calculate the Berry curvature dipole in the presence of strain. We show that the strained tWSe$_2$ exhibits large nonlinear Hall response. Third, we explore the behavior of the NLH response near the topological phase transition induced by a displacement field. We find that the Berry curvature dipole is strongly enhanced near the topological phase transition point and it changes sign across the transition point.

\begin{figure}
	\centering
	\includegraphics[width=1\linewidth]{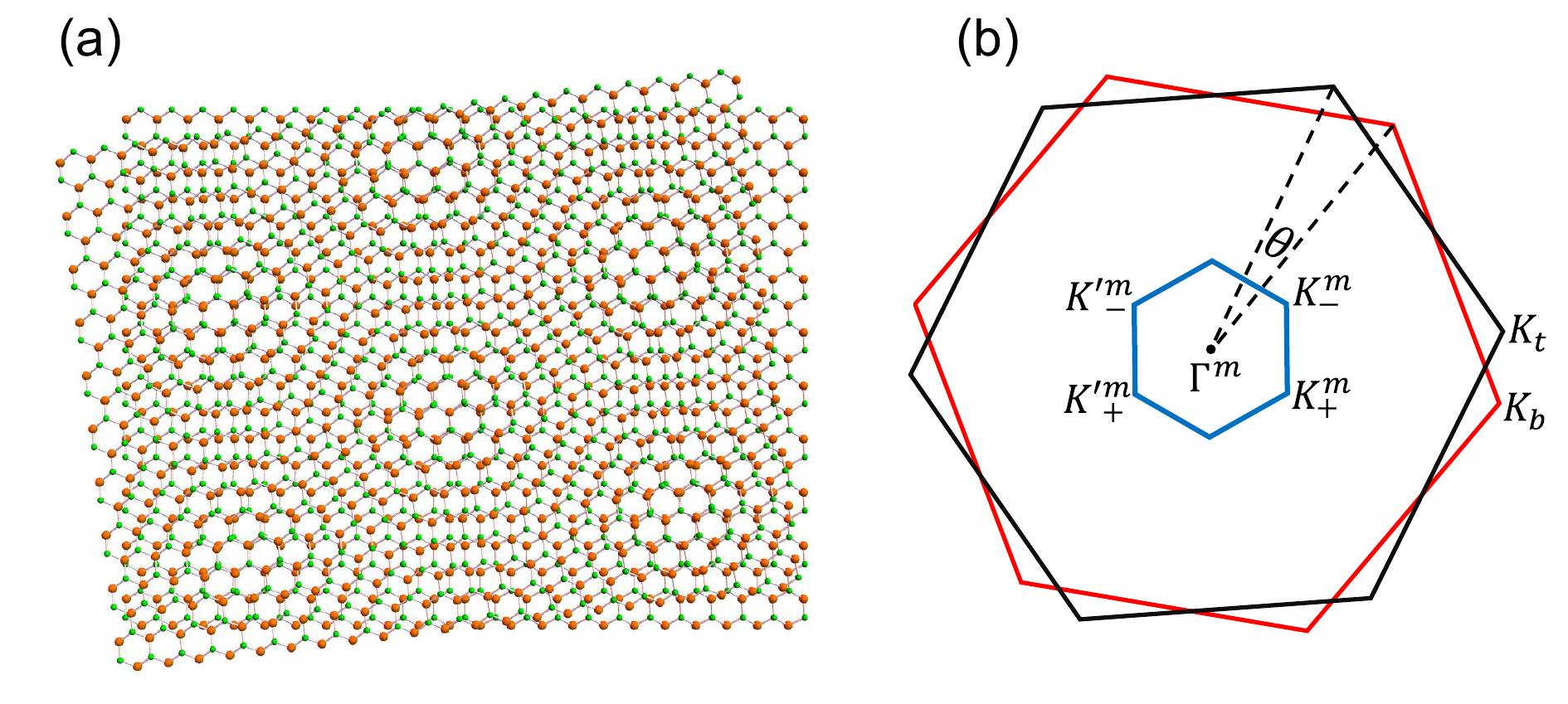}
	\caption{Schematic figure and Brillouin zone of tWSe$_2$. (a) A schematic figure of tWSe$_2$ moir\'{e} superlattice at a twist angle of $7^\circ$. (b) The original Brillouin zone of bottom (red) and top (black) layers and the folded moir\'{e} Brillouin zone (blue). $\theta$ is the twist angle between the two layers. $K_b$, $K_t$, $K_{\pm}^m$, $K_{\pm}^{'m}$ are moir\'{e} Brillouin zone corners. $\Gamma^m$ is the center of moir\'{e} Brillouin zone.}
	\label{fig1}
\end{figure}

\emph{Continuum model of strained tWSe$_{2}$.}---We consider a AA stacking bilayer WSe$_{2}$ with lattice constant $a_0$ in a single layer and at a twist angle $\theta$. A schematic figure is shown in Fig.~\ref{fig1}a. The moir\'{e} superlattice, which has a moir\'{e} lattice constant $L_M=a_0/2\sin\frac{\theta}{2}$, folds the energy bands and  gives rise to the so-called moir\'{e} Brillouin zone (see Fig.~\ref{fig1}b). These moir\'{e} energy bands originating from the $K_{+}$ or $K_{-}$ valleys can be described by the continuum Hamiltonian\cite{wu2019topological} $H=\sum_{\xi}\int d\bm{r}\psi_{\xi}^{\dagger}(\bm{r})\mathcal{\hat{H}}_{\xi}(\bm{r})\psi_{\xi}(\bm{r})$ with
\begin{equation}
\hat{H}_{\xi}(\bm{r})=\begin{pmatrix}
\hat{H}_{b,\xi}+\Delta_b(\bm{r})& T_{\xi}(\bm{r})\\
T_{\xi}^{\dagger}(\bm{r})&\hat{H}_{t,\xi}+\Delta_t(\bm{r})
\end{pmatrix}
\label{eq:continumm}.
\end{equation}
Here, $\xi = \pm$ is the valley index denoting whether the bands originate from the $K_{+}$ or $K_{-}$ valleys of the monolayer Brillouin zone. The electron creation operators are denoted as $\psi_{\xi}^{\dagger}=(\psi^{\dagger}_{b,\xi},\psi^{\dagger}_{t,\xi})$, where $t$ and $b$ label the top and bottom layers respectively. It is important to note that due to the large Ising spin-orbit coupling in 2H-structure WSe$_2$, the top valence band of monolayer WSe$_2$ $K_{+}$ and $K_{-}$ valleys are fully spin polarized and have opposite spin. Therefore, the spin and valley indices are locked together and the spin-index is dropped. As a result, the Hamiltonian of a single layer at valley $\xi$ can be written as
\begin{equation}
\hat{H}_{l,\xi}=-\frac{\hbar^2}{2m^*} (\hat{\bm{k}}-\bm{K}^{m}_{l,\xi})^2-l\frac{V_z}{2},
\end{equation}
where   $l=+1(-1)$ labels the bottom(top) layer, $m^*$ is the effective mass of valence band, and  $V_z$ is the staggered layer potential generated by the vertical displacement field. The intra-layer moir\'{e} potential and the coupling between the two layers are denoted as $\Delta_{l}(\bm{r})$ and $T_{\xi}(\bm{r})$ respectively which can be written as:
\begin{eqnarray}
\Delta_{l}(\bm{r})&=&V\sum_{i=1,2,3}e^{i(\bm{g_i}\cdot\bm{r}+l\psi)}+h.c.,\\
T_{\xi}(\bm{r})&=&w(1+e^{-i\xi\bm{g_2}\cdot\bm{r}}+e^{-i\xi (\bm{g_1}+\bm{g_2})\cdot\bm{r}}).
\end{eqnarray}
Here,  $V$ and $\psi$, characterize the amplitude and phase of the moire potentials respectively, $w$ characterizes the tunneling strength between the top and bottom layers, and the moir\'{e} reciprocal lattice vectors are $\bm{g_{i}}=\frac{4\pi}{\sqrt{3}L_M}(\cos\frac{2(i-1)\pi}{3},\sin \frac{2(i-1)\pi}{3})$. We adopt the model parameters from  ref.~\cite{wu2019topological}$(a_0,m^*,w,V,\psi)=(3.32\AA,0.44m_e,9.7 \text{ meV},8.9 \text{ meV},91^\circ)$, which are estimated from the first principle calculations.

\begin{figure}
	\centering
	\includegraphics[width=1\linewidth]{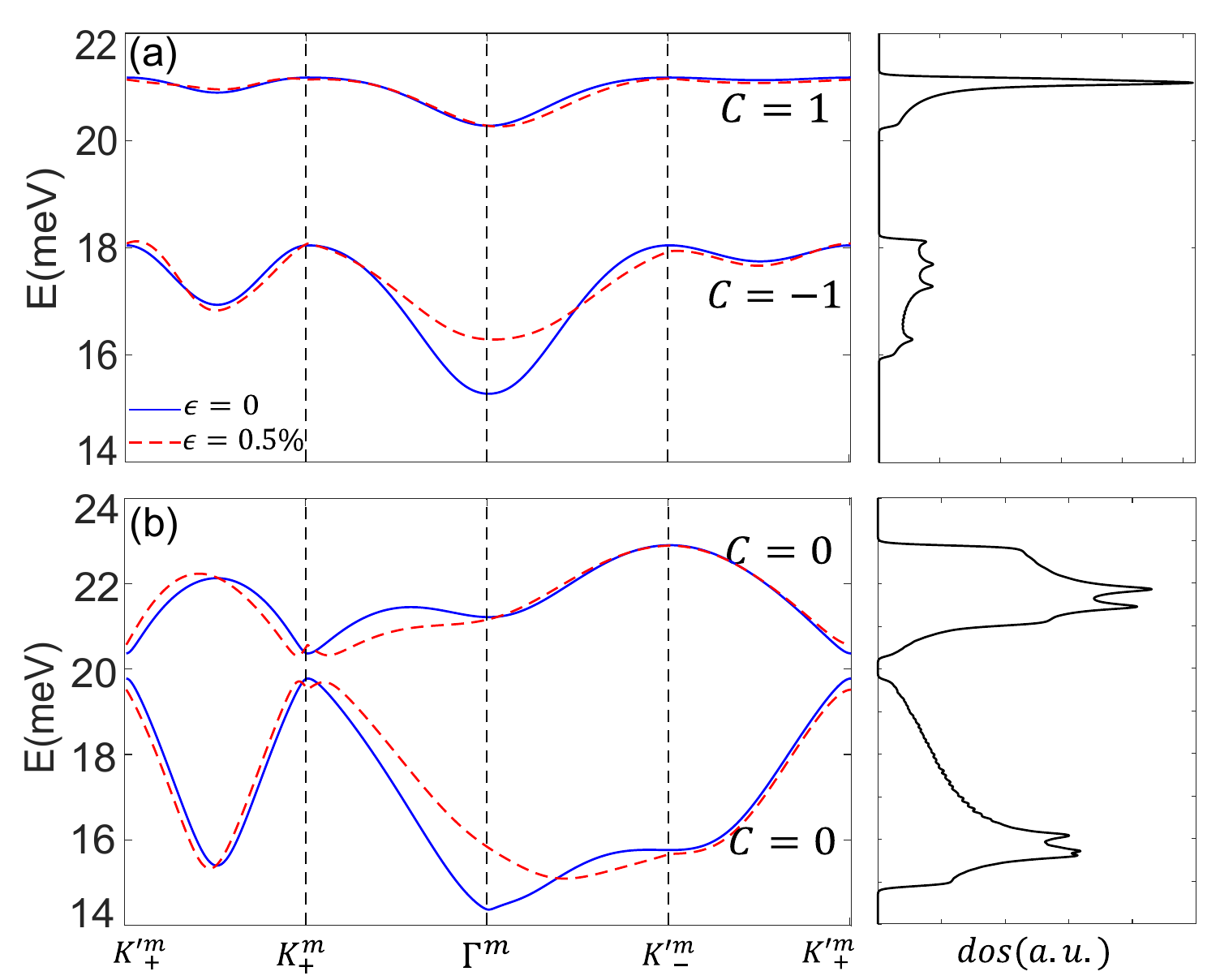}
	\caption{The band structure and density of states (dos) of tWSe$_2$. (a) The band structure of tWSe$_2$ at $\theta=1.4^\circ$ with (dashed red lines) and without (solid blue lines) strain. The staggered potential $V_z=0$ and the top two valence bands originating from $K_+$ valleys of the original monolayer Brillouin zone have opposite Chern numbers. The bands originating from the $K_-$ valleys are related to the $K_+$ valley bands by time-reversal which are not shown above. (b) The band structure with $V_z=5$ meV with (dashed red lines) strain and without (solid blue lines) strain. As $V_z$ increases, a topological phase transition happens at $V_z \approx 4.2$ meV and the two top valence bands exchange Chern numbers and results in topologically trivial flat bands. }.
	\label{fig2}
\end{figure}

It is important to note that the Berry curvature dipole transforms as a pseudo-vector which vanishes in the presence of a three-fold rotational symmetry as discussed in detail in the next section. In realistic systems, the three-fold symmetry in tWSe$_2$ can be broken by the substrate induced or externally applied strain. The strain induced symmetry breaking has been observed experimentally\cite{zhang2020flat}. In general, the physical properties of tWSe$_2$ only depend on the relative deformation between the two layers. Therefore for all the calculations in this work, the strain is only applied to the bottom layer of WSe$_2$, while the strain in the top layer is set to zero. Specifically, the strain tensor $ \bm{\mathcal{E}}$ is a two-dimensional matrix, which can be written as\cite{bi2019designing}:
\begin{equation}
\bm{\mathcal{E}}=\epsilon\left(
\begin{array}{cc}
\cos^2\varphi-\nu\sin^2\varphi & (1+\nu)\cos\varphi\sin\varphi\\
(1+\nu)\cos\varphi\sin\varphi  & \sin^2\varphi-\nu\cos^2\varphi\\
\end{array}
\right).
\end{equation}
The angle $\varphi$ denotes the in-plane direction of the uniaxial strain direction with respect to the zigzag edge of the sample and $\epsilon$ characterizing the strength of strain, and $\nu=0.19$ is the poisson ratio for WSe$_2$\cite{kang2013band}.

In general, the strength $\epsilon$ can be estimated to $0.5\%\sim 1\%$, but the direction $\varphi$ can be random and difficult to determine. We highlight the breaking of the $C_3$ rotational symmetry due to strain is the key of the nonlinear Hall response, while the details of the strain are not essential for our discussion. For simplicity, we use the model that there is no strain in the top layer and the strain is applied to the bottom layer.The strain has two important effects shifting the Dirac point for bottom layer WSe2 to $\bm{D}_{b,\xi}=(I-\bm{\mathcal{E}}^T)\bm{K}_{b,\xi}-\xi\bm{A}$ and generating an effective gauge field $\bm{A}=\frac{\sqrt{3}}{2a_0}\beta(\epsilon_{xx}-\epsilon_{yy},-2\epsilon_{xy})$. Here, $\beta$ is taken as 2.30 in our calculations according to previous first principle calculations for strained monolayer WSe$_2$ \cite{fang2018electronic,bi2019designing}. As a result, the continuum Hamiltonian of this strained tWSe$_2$  is obtained by replacing $\hat{H}_b$ in Eq.~\ref{eq:continumm} as
\begin{equation}
\hat{H}_{b,\xi}^s=-\frac{\hbar^2}{2m^*} ((I+\bm{\mathcal{E}}^T)(\hat{\bm{k}}-\bm{D}^{m}_{b,\xi}))^2-\frac{V_z}{2}.
\end{equation}

To estimate the range of strain $\epsilon$ within which our model is valid, we expect that the strain induced shifting of the Dirac point would be smaller than the separation of the Dirac points due to twisting. With a twist angle $\theta$, the separation of the Dirac point $K$ in momentum space is $\Delta K=|K|\theta$, and the shift of the $K$ point by uniaxial strain is $\Delta K_s\approx \epsilon|K|$. Therefore,  the strain effect can be treated as a perturbation to the moir\'{e} super-lattice when $\epsilon \ll \theta$.  In our calculations below, we assume that the strain induces a $0.5\%$ change in the lattice constant of the bottom layer WSe2 along the direction of the strain such that  $\epsilon \approx 0.2\theta$.

In Fig.~\ref{fig2}a, we show the moir\'{e} energy bands of tWSe$_2$ with twist angle $\theta=1.4^{\circ}$ with and without strain. The top two valence bands originating from the $K_{+}$ valley of the monolayer Brillouin zones carry finite Chern numbers $C=1$ and $C=-1$ respectively when $V_z=0$. Due to time-reversal symmetry, the top two valence bands originating from the $K_{-}$ valleys carry Chern numbers $C=-1$ and $C=1$ respectively. As the two valleys do not couple in momentum space, one can define a $Z_2$ topological invariant as $Z_2=|(C_{K_{+}}-C_{K_{-}})/2|$ to describe the topological properties of the bulk bands, where $C_{K}$ is the Chern number of the top valence band originating from the $K$ valleys of the monolayer Brillouin zone. In Fig.~\ref{fig2}b, a staggered potential of $V_z=5$ meV induces a topological phase transition and the system becomes topologically trivial. The band structure with and without strain are depicted. It is important to note that the strain with $\epsilon = 0.5\%$ is not sufficient to close the band gap and the topological properties of the flat bands are not changed by strain. Moreover, the energy dependence of the density of states for $V_z=0,5$ meV are also shown on the right side of Fig.~\ref{fig2}, which is important to determine the occupation number of holes $n_h$ as discussed below.

\emph{Nonlinear Hall response.}---In this section, we consider the NLH response for tWSe$_2$. The NLH effect is characterized by the generation of a transverse voltage using a charge current in time-reversal invariant systems without external magnetic fields or magnetic orders. Moreover, this effect is nonlinear in nature and exhibits a quadratic current-voltage relation. More specifically, when an electric field $\bm{E}(t)=\frac{1}{2}(\varepsilon e^{i\omega t}+\varepsilon^*e^{-i\omega t})$ with the amplitude vector $\varepsilon$ and frequency $\omega$ is applied, the Hall current has both rectified and second-harmonic components  $J_{y(x)}^{(0)}=\chi_{yxx(xyy)}\varepsilon_{x(y)}\varepsilon_{x(y)}^*$ and $J_{y(x)}^{(2)}=\chi_{yxx(xyy)}\varepsilon_{x(y)}\varepsilon_{x(y)}$, where $\chi$ is the nonlinear Hall susceptibility. As shown in ref.~\cite{sodemann2015quantum,zhou2020highly,du2018band}, the nonlinear Hall susceptibility can be written as:
\begin{equation}
\chi_{yxx(xyy)}=\mp\frac{e^3\tau}{2\hbar^2(1+i\omega\tau)}D_{x(y)}
\end{equation}
where $\tau$ is the relaxation time. The Berry curvature dipole $D_{i}$ are elements of the Berry curvature dipole pseudo-vector. It is important to note that in a two-dimensional system, the Berry curvature dipole transforms as a pseudo-vector as described by the vector $D_{i}$ ($i=x,y$). In other words, $D_{i}$ must be invariant under crystal point group operations. For twisted tWSe$_2$ with or without the displacement field, the system respects a three-fold rotational symmetry $C_{3z}$ inherited from monolayer 2H-structure WSe2. Due to the time-reversal symmetry, $\Omega_n(\bm{k}+\bm{K})=-\Omega_n(-\bm{k}-\bm{K}), \bm{v}_n(\bm{k}+\bm{K})=-\bm{v}_n(-\bm{k}-\bm{K})$, each valley has an equal contribution, allowing us to consider the $K_{+}$ valley for the sake of simplicity. For $K_{+}$ valley, $C_{3z}$ symmetry ensures $\Omega_n(\bm{k})=\Omega_n(C_3\bm{k})=\Omega_n(C_3^2\bm{k})$ and $\sum_{i=0}^2 \bm{v}_n(C_3^i\bm{k})=0$.  Therefore, $D_i$ vanishes if $C_{3z}$ symmetry is preserved. One way to obtain finite $D_{i}$ is to take into account the strain effects which break $C_{3z}$. The strain can be induced by the substrate which couples to tWSe$_2$ or it can be induced externally as shown in the above sections. As we will see in the next section, only a very small strain is needed to induce a strong NLH response in tWSe$_2$.

\emph{Topological phase transition and NLH effect.}---With the formalism discussed above, we can now explore the NLH effects in tWSe$_2$. By applying a unaxial strain with $\epsilon=0.5\%$ along the direction of the zigzag edge and at twist angle $1.4^\circ$, $D_x$ as a function of the vertical displacement
 field $V_z$ and the occupation number of holes per moire unit cell $n_h$ are depicted in Fig.~\ref{fig3}a. The occupation number $n_h$ is obtained by integrating the density of states from the top of the valence to the Fermi energy including two valleys. It is clear from Fig.~\ref{fig3}a that the Berry curvature dipole is generally large, in the order of $1\AA$. %This is comparable to the Berry curvature dipole found in previous experimental systems\cite{QMa,KaifeiKang}.

\begin{figure}
	\centering
	\includegraphics[width=1.0\linewidth]{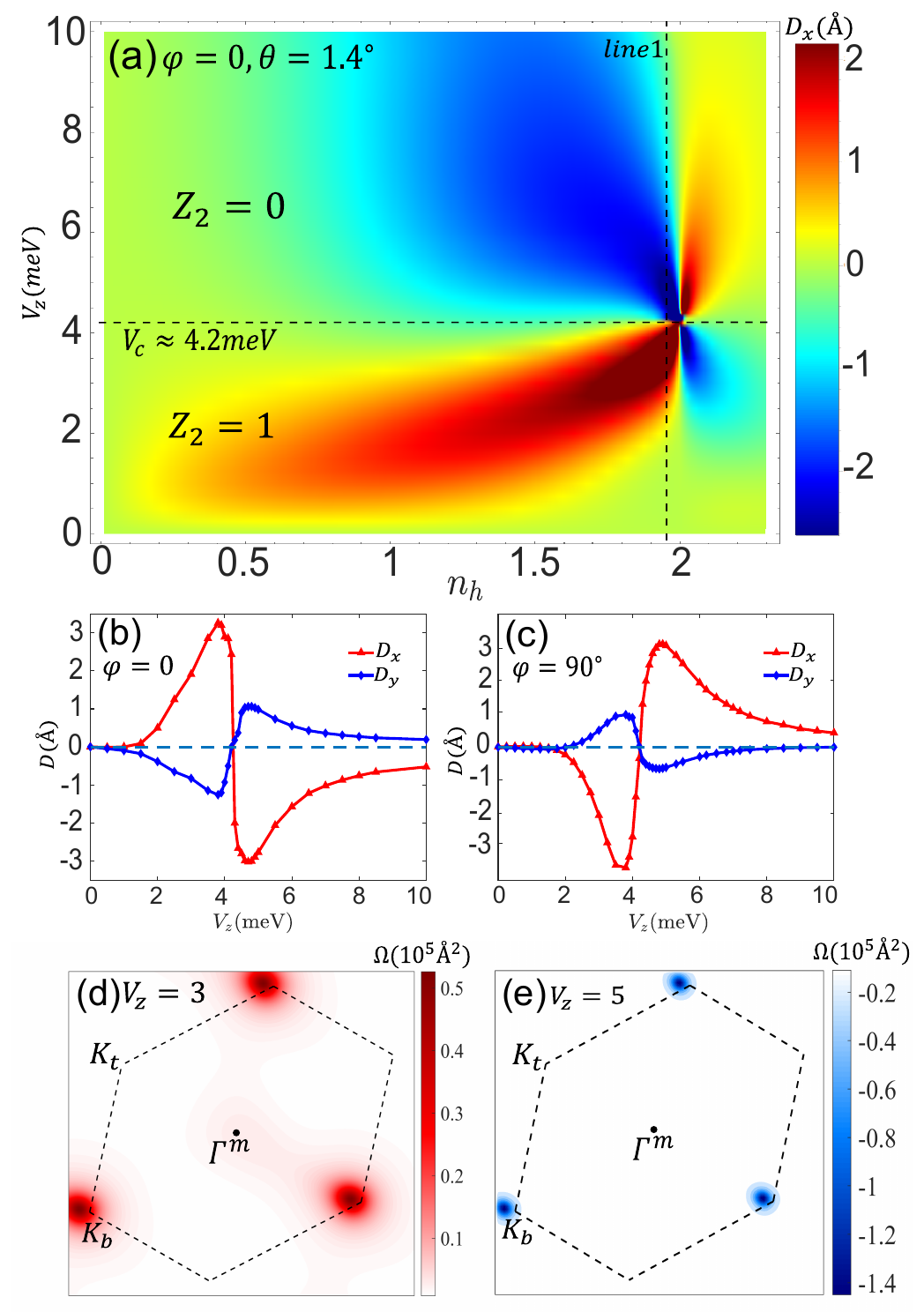}
	\caption{Berry curvature dipole with different strain and displacement fields. (a) The $V_z$(out of plane displacement field) and $n_h$(number of holes per unit cell) dependence of $D_{x}$($x$ component Berry dipole) with strain $\epsilon=0.5\%$ along the zigzag edge direction. The color bar shows the magnitude of $D_x$. The critical $V_z$ at which the phase transition occurs is denoted by the horizontal dashed line. $D_x$ is strongly enhanced near the phase transition point and $n_h \approx 2$ when the Fermi energy is near the bottom of the top valence band. (b) $D_{x}, D_{y}$ as a function of $V_z$ with fixed $n_{h}$. The values of $n_h$ ($=1.95$) are denoted by the vertical line 1 in (a). For (b), an uniaxial strain is applied along the zigzag edge direction. (c) Same parameters as (b) but the uniaxial strain is applied along the armchair edge direction. (d) and (e) The Berry curvatures $\Omega$ of top valence band of strained tWSe$_2$ in the deformed Brillouin zone before and after the topological phase transition respectively. In (d), $V_z = 3$ meV, in (e), $V_z = 5$ meV. The strain is applied along the zigzag edge direction. The temperature is set to $1.5 K$.}.
	\label{fig3}
\end{figure}

Importantly, the Berry curvature dipole is strongly enhanced near the bottom of the top valence band when $n_h$ is close to 2 (this happens when the Fermi energy is near the bottom of the top valence band) and $V_z$ close to 4.2 meV. Specifically, the bands originating from the $K_{+}$ (or $K_{-}$) valley have nontrivial topological invariant $Z_2=1$. By applying a displacement field which introduces a potential difference between the top and the bottom layers, the top valence band and the nearby valence band can touch near the $K$ points and exchange Berry curvatures, resulting in a change of $Z_2$ invariant from 1 to 0. The $V_z$ dependence of the Berry curvature dipole across the phase transition with fixed $n_h$ is shown in Fig.~\ref{fig3}b. It is clear that Berry curvature dipole is strongly enhanced near the critical $V_z \approx 4.2$ meV when $n_h$ is close to 2. The enhancement of Berry curvature dipole comes from two contributions, one is the enhancement of density of states due to the band flatness, and the other is the large Berry curvature near band edge. This value is several times larger than the optimal Berry curvature dipole measured experimentally in other systems which is about $1\AA$ \cite{ma2019observation,kang2019nonlinear,zhang2018electrically}. Moreover, the Berry curvature dipole changes sign across the phase transition point. This is caused by the exchange of Berry curvatures near the $K$ points before and after the topological phase transition as shown in Fig.~\ref{fig3}d and Fig.~\ref{fig3}e. To illustrate the sign change of Berry dipole more clearly, we show the Berry dipole flow $v_i\Omega$ in Supplementary note 1.

Furthermore, the $V_z$ dependence of the Berry curvature dipole with fixed $n_h$ is shown in Fig.~\ref{fig3}c when strain is applied along the armchair edge direction. Similar enhancement and sign change in the Berry curvature dipole near the topological phases transition can be observed. In the supplementary note 2, the $V_z$ and $n_h$ dependence of the Berry curvature dipole with different twist angles and strain are presented and behaviors similar to Fig.~\ref{fig3}a are observed. Therefore, the NLH measurements can be used to indicate possible topological phase transitions in tWSe$_2$. Note that in practice to confirm the sign switching of NLH signal is induced by the topological phase transition rather than other mechanisms, it must be accompanied by other measurements that can probe the bulk topology or edge excitations.

\emph{Discussion.}---It is important to note that the appearance of NLH effect studied in this work is very general. We expect a finite NLH response as long as the three-fold rotational symmetry is broken intrinsically by the substrate\cite{zhang2020flat} or by an externally applied strain. The calculated Berry dipole in tWSe$_2$ is about $1\sim 3\AA$, which is relatively large in usual materials. In other strained systems, such as strained MoS2 and strained bilayer graphene\cite{son2019strain,battilomo2019berry}, the strain-induced Berry dipole is only $10^{-2}\AA$. For bilayer and few-layer WTe$_2$\cite{du2018band,wang2019ferroelectric}, the Berry dipole is about $0.2\AA,1\AA$ respectively.

Furthermore, we point out that strained tWSe$_2$ and other moir\'e materials are excellent candidate materials for studying NLH effects due to their low symmetry, non-trivial Berry curvature and high tunability of experimental parameters\cite{chen2020tunable,zhang2019nearly}. Especially, the $C_{3z}$ symmetry breaking strain have been widely observed in moir\'e materials. In our other recent work collaborating with experimentalists\cite{huang2022giant},  it was found that the twisted monolayer-monolayer, bilayer-bilayer and trilayer-trilayer WSe$_2$ samples, all of them exhibit strong nonlinear hall response, roughly 1000 times larger than those of bilayer or few-layer WTe$_2$\cite{ma2019observation,kang2019nonlinear}. This clearly shows that the moir\'e superlattices behave as an ideal setup in realizing the nonlinear Hall effects, although the understanding of such a giant nonlinear hall response in the moir\'e superlattices is still open and worth further works in the future.

Our calculations are based on the Boltzmann approach in Ref.~\cite{sodemann2015quantum}, which requires that the inelastic scattering time $\tau_{in}$ and the band width $W$ should satisfy $W\gg \hbar/\tau_{in}$. Although the bandwidth is reduced at orders of meV by the moir\'e  potential in tWSe$_2$, we assume the Boltzmann approach is still valid in this case.  This assumption is based on the observation that $\tau_{in}$ is usually around 10 ps for 2-D van der Waals materials\cite{wu2007weak}, corresponding to an energy scale $\hbar/\tau_{in}\sim 0.07$ meV that is still much smaller than the moir\'e band width.

The electron-electron interaction in general is expected to play an important role in moir\'e superlattices. Recently, correlated insulating phase is observed when the filling factor $n_h$ is close to 1\cite{wang2020correlated}. In principle, the interaction can modify the single-particle band structures and further  change the Berry curvature dipole.  In the experiment\cite{huang2022giant}, the nonlinear Hall response is strongly enhanced at half filling of the first moir\'e band, which cannot be explained in single-particle level and correlation effects must have contributions to the NLH response. Near half-filling of the first moir\'{e} band, the
nonlinear Hall signal shows a sharp peak which can originate from a mass-diverging type continuous Mott transition. Also the symmetry-breaking ground states as discussed in \cite{po2018origin} may also get involved and interplay with NLH. For example, the $C_{3z}$ rotational symmetry can be broken in the presence of the nematic phase, which is essential in generating a finite NLH in tWSe$_2$. The interplay between the interaction effects and nonlinear Hall effect is of great interest and is worth a more detail study in the future.

In addition, the topological phase transition happens near $n_h \approx 2$. For a fixed displacement field $V_z$, the Berry curvatures near the band edge of top two valence bands are opposite and as a result, the Berry dipoles change sign when filling factor $n_h$ across 2. For a fixed $n_h$, by applying $V_z$  which introduces a potential difference between the top and the bottom layers, the top valence band and the nearby valence band can touch near the $K$ points and exchange Berry curvatures, resulting in the sign change of $D_i$. This is the origin of four wings of the "butterfly" in Fig.~\ref{fig3}a. Near the phase transition regime, we expect the system to be metallic and can be described by a Fermi liquid even in the presence of electron-electron interaction. Nevertheless, the NLH measurements can reveal the topological properties of the bands away from the Fermi energy that shed light on the nature of correlated phases.

\emph{Methods.}---In strictly two dimensions, the Berry curvature dipole is given by
\begin{equation}
D_{i}=-\int \frac{d\bm{k}}{(2\pi)^2}\sum_{n,\xi=\pm1}v_{n\bm{k},\xi}^i\Omega_{n\bm{k},\xi} \frac{\partial f(E_{n\bm{k},\xi})}{\partial E}, \label{Berry_dipole}
\end{equation}
where $n$ as the band index and $v^{i}$ as the Fermi velocity of the Bloch state, and the Berry curvature can be obtained from Bloch wavefunctions as:
\begin{equation}
\Omega_{n\bm{k},\xi}=i\langle \partial_{\bm{k}}u_{n\bm{k},\xi}|\times |\partial_{\bm{k}}u_{n\bm{k},\xi}\rangle .
\end{equation}
The Berry curvature dipoles should be integrated over whole moir\'{e} Brillouin zone and both valleys are involved in the result. Also the
Chern number for an isolated band can be evaluated from Berry curvature:
\begin{equation}
C_{n\xi}=\frac{1}{2\pi}\int d\bm{k}\Omega_{n\bm{k},\xi}
\end{equation}

\emph{Acknowledgments.}---We are grateful to the illuminating discussions with Meizhen Huang, Zeifei Wu and Ning Wang. We thank the support of the Croucher Foundation and HKRGC through 16324216, 16307117 and 16309718.

\clearpage
		\onecolumngrid
\begin{center}
		\textbf{\large Supplementary Material for\\ ``Nonlinear Hall Effects in Strained Twisted Bilayer WSe$_2$''}\\[.2cm]		
      Jin-Xin Hu,$^{1}$  Cheng-Ping Zhang,$^{1}$  Ying-Ming Xie,$^{1}$  K. T. Law$^{1}$\\[.1cm]
		{\itshape ${}^1$Department of Physics, Hong Kong University of Science and Technology, Clear Water Water Bay,  Hong Kong, China}
\end{center}
	
	\maketitle

\setcounter{equation}{0}
\setcounter{section}{0}
\setcounter{figure}{0}
\setcounter{table}{0}
\setcounter{page}{1}
\renewcommand{\theequation}{S\arabic{equation}}
\renewcommand{\thesection}{ \Roman{section}}

\renewcommand{\thefigure}{S\arabic{figure}}
\renewcommand{\thetable}{\arabic{table}}
\renewcommand{\tablename}{Supplementary Table}

\renewcommand{\bibnumfmt}[1]{[S#1]}
\renewcommand{\citenumfont}[1]{#1}
\makeatletter

\maketitle

\setcounter{equation}{0}
\setcounter{section}{0}
\setcounter{figure}{0}
\setcounter{table}{0}
\setcounter{page}{1}
\renewcommand{\theequation}{S\arabic{equation}}
\renewcommand{\thesection}{ \Roman{section}}

\renewcommand{\thefigure}{S\arabic{figure}}
\renewcommand{\thetable}{\arabic{table}}
\renewcommand{\tablename}{Supplementary Table}

\renewcommand{\bibnumfmt}[1]{[S#1]}
\makeatletter

\maketitle

\section*{\bf{\uppercase\expandafter{SUPPLEMENTARY NOTE 1: Berry dipole density near topological phase transition}}}
\begin{figure}[h]
	\centering
	\includegraphics[width=1.0\linewidth]{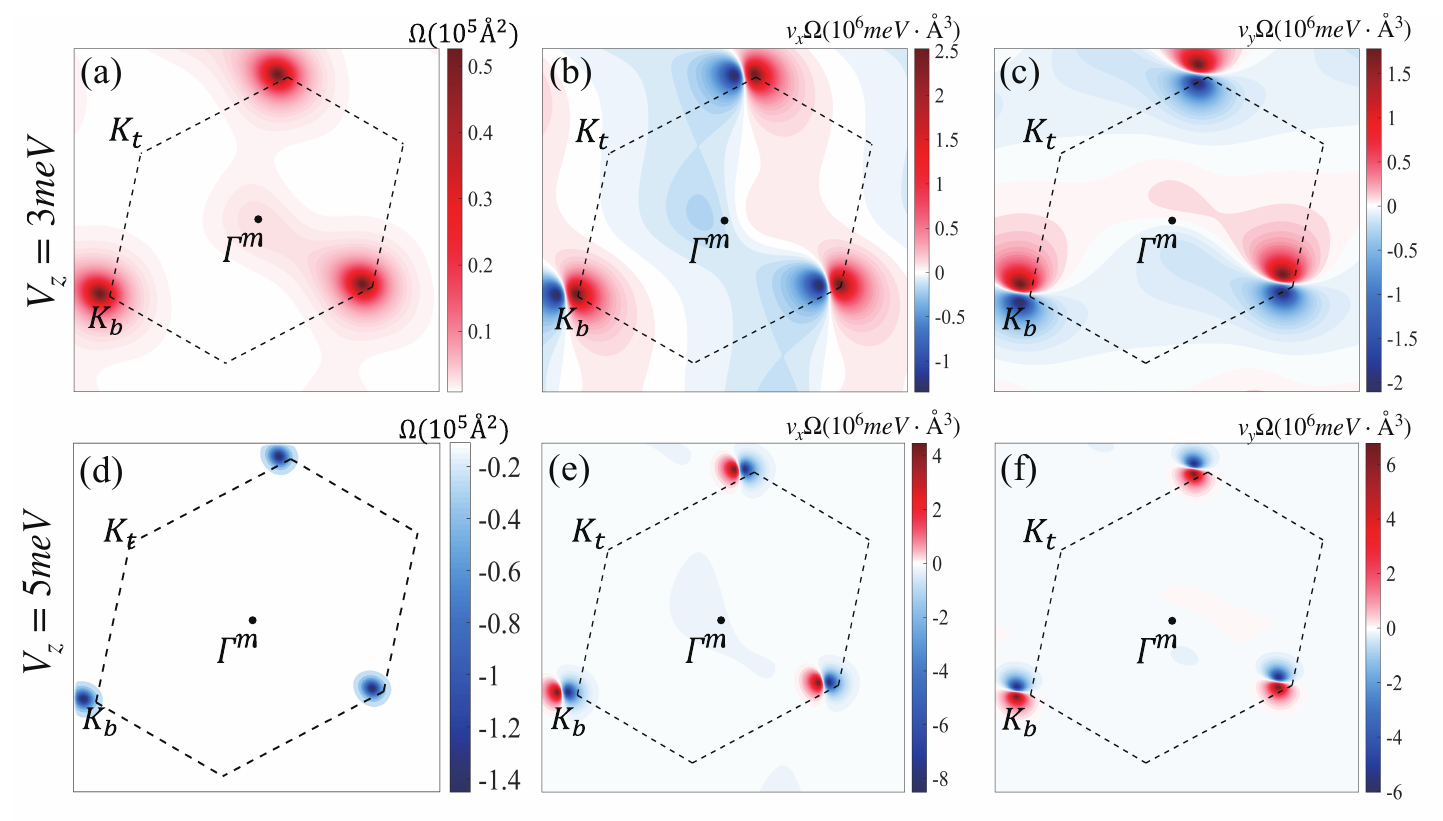}
	\caption{(a)-(c) The Berry curvature $\Omega$, Berry-dipole density $v_x\Omega$ and $v_y\Omega$ for $V_z=3meV$. (d)-(f) The Berry curvature $\Omega$, Berry-dipole density $v_x\Omega$ and $v_y\Omega$ for $V_z=5meV$. $\theta=1.4^\circ$ and the strain $\epsilon=0.5\%$ along the zigzag edge direction.}.
	\label{FIGS1}
\end{figure}

In Fig.~\ref{FIGS1} we plot the Berry curvature $\Omega$ as well as Berry dipole density $v_i\Omega$ for $\theta=1.4^\circ$. It is shown clearly that the Berry-dipole densities near Brillouin zone corner $K_b$ contribute most to the Berry dipole. When $V_z$ crosses the critical field $V_c\approx 4.2meV$ during the topological phase transition, the dipole densities change sign, which give the sign change of nonlinear Hall signal.

\section*{\bf{\uppercase\expandafter{SUPPLEMENTARY NOTE 2: Berry curvature dipole for different twist angles and Strain}}}
\begin{figure}[htbp]
	\centering
	\includegraphics[width=0.6\linewidth]{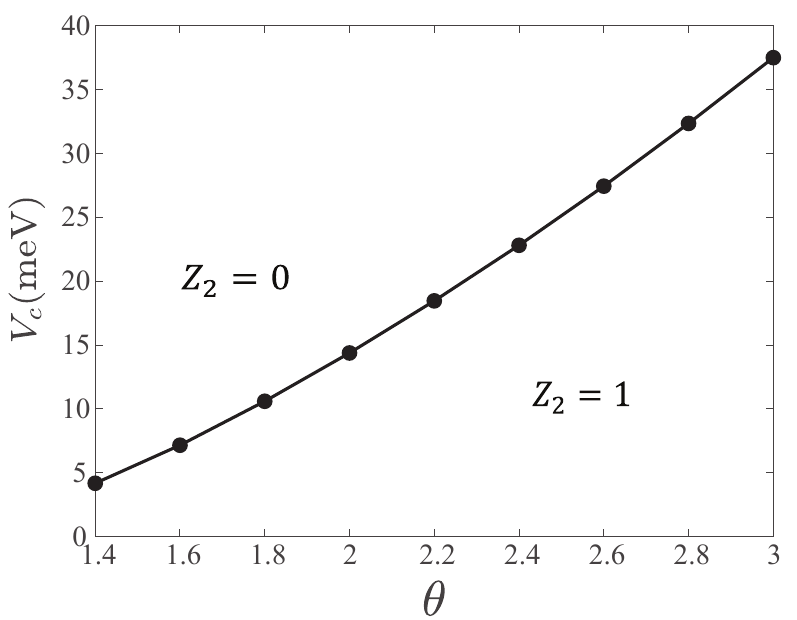}
	\caption{the topological phase transition critical layer potential $V_c$ as a function of twist angle $\theta$}.
	\label{FIGS2}
\end{figure}

In Fig.~\ref{FIGS2} we show the critical layer potential $V_c$ as a function of twist angle $\theta$ in the range of $1.4^\circ \sim 3^\circ$. It is clear that the $V_c$ increases with $\theta$. We estimate the relationship between displacement field $D$ and layer potential $V_z$ by
\begin{equation}
  V_z=eDd_s/\varepsilon^r
\end{equation}
Here $d_s\approx 0.5nm$ is the spatial separation between top and bottom layers\cite{wu2019topological}, and $\varepsilon^r$ is relative dielectric constant. For most of devices, the dielectric layers are composed of SiO$_2$ and hBN, thus $\varepsilon^r\approx 4$\cite{shen2020correlated}. According to Eq.S1, for $\theta = 1.4^\circ \sim 3^\circ$, the critical displacement field $D_c=0.03 \sim 0.3 V/nm$, which is a region that can be reached experimentally\cite{shen2020correlated}.

\begin{figure}[h]
	\centering
	\includegraphics[width=1.0\linewidth]{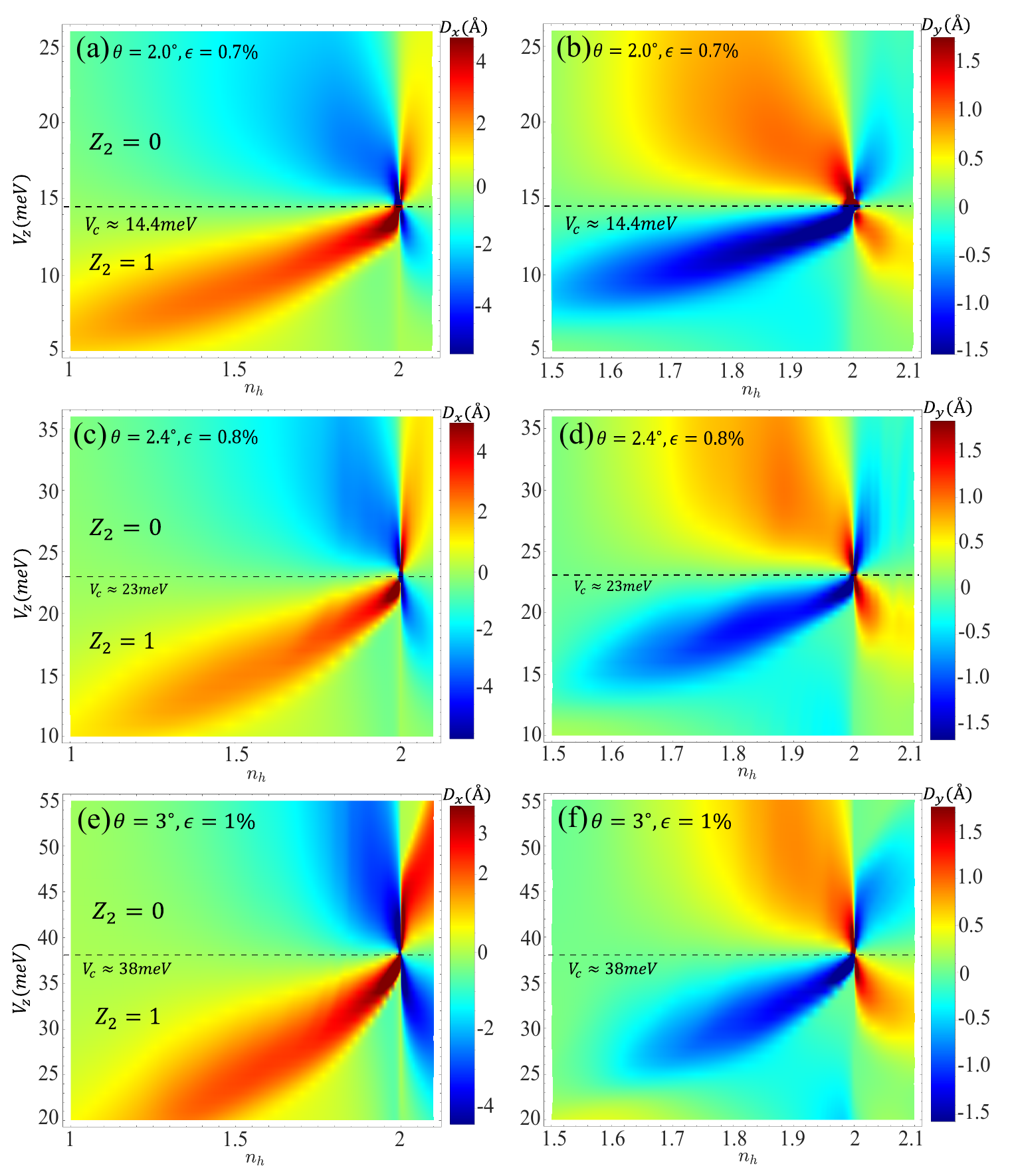}
	\caption{The $V_z$ and $n_h$ dependence of the Berry curvature dipole $D_x$ and $D_y$ at $\theta=2^\circ, \epsilon=0.7\%$ for (a)(b), $\theta=2.4^\circ, \epsilon=0.8\%$ for (c)(d) and $\theta=3^\circ, \epsilon=1\%$ for (e)(f). The strain direction is along the zigzag edge.}.
    \label{FIGS3}
\end{figure}

To make a supplement, we present the calculation of Berry curvature dipole for $\theta=2.0^\circ$, $2.4^\circ$ and $3.0^\circ$. Fig.~\ref{FIGS3} shows the $x$ and $y$ components of berry curvature dipole for different twist angles and strain strength. The strain strength is chosen according to the approximation in main text which satisfies $\epsilon =0.2\theta$. It is clear shown that the Berry curvature dipole is strongly enhanced near $n_h\approx 2$ and $V_c$, which is similar to Fig.3(a) in main text.
\end{document}